# Polymer/2D material nanocomposite manufacturing beyond laboratory frontiers

Pablo A.R. Muñoz[1], Camila F. P. de Oliveira[1], Leice G. Amurin[1], Camila L.C. Rodriguez[1], Danilo A. Nagaoka[1], Maria Inês Bruno Tavares[2], Sergio H. Domingues[1], Ricardo J. E. Andrade[1], Guilhermino J. M. Fechine*[1]

*Polymer nanocomposites based on 2D materials as fillers are the target in the industrial sector, but the ability to manufacture them on a large scale is very limited, and there is a lack of tools to scale up the manufacturing process of these nanocomposites. Here, for the first time, a systematic and fundamental study showing how 2D materials are inserted into the polymeric matrix in order to obtain nanocomposites using conventional and industrially scalable polymer processing machines leading to large-scale manufacturing are described. Two new strategies were used to insert pre-exfoliated 2D material into the polymer matrix, liquid-phase feeder, and solid-solid deposition. Characterizations were beyond micro and nanoscale, allowing the evaluation of the morphology for millimeter samples size. The methodologies described here are extendable to all thermoplastic polymers and 2D materials providing nanocomposites with suitable morphology to obtain singular properties and also triggering the start of the manufacturing process on a large scale.*

In 2004, a single-crystalline graphite film of atomic thickness[1] was isolated. The researchers responsible for this revolutionary work, K. S. Novoselov and A. Geim, were shortly thereafter awarded the Nobel Prize in Physics for "innovative experiments with graphene" in 2010. Several researchers have used graphene (Gr), graphene oxide (GO), and reduced graphene oxide (rGO) as mechanical reinforcing nanoparticles in polymeric matrices. The insertion of these two-dimensional (2D) materials considerably improves the mechanical properties of the polymers. They also increase the thermal and electrical conductivity and the dimensional stability of the composite when compared to the polymer matrix[2,3]. Important challenges still need to be overcome to produce polymer nanocomposites based on two-dimensional particles (graphene-based materials, molybdenum disulfide, hexagonal boron nitride, and phosphorene among others) on a large scale. Essentially, there are three strategies for nanocomposites preparation: 1) Solution mixing, 2) In situ polymerization, and 3) Melt mixing. The first two options achieve excellent results in terms of particles dispersion, however, the scale-up of these methods to industrial production is limited. Melt mixing is the only one that can be easily scaled-up using equipment such as a twin screw extruder machine as part of a plastic production line.

The most important advantage of two-dimensional (2D) particles is their large surface area that enhances the interface with the polymer. This peculiarity leads to significant improvements in different properties of the polymer matrix with the insertion of very small contents of particles, which may not reach 2% by mass[4–7]. However, if the dispersion is not efficient, the re-stacking of the 2D material sheets occurs in excessive amounts, and the above-mentioned advantage is lost. In the case of two-dimensional particles, the high dispersion degree in the polymeric matrix by using melt mixing process might turn in a challenge due to the need for tuning the processing conditions (screw speed, residence time of the 2D particles, processing temperature, and shear stress) for the singular shape of 2D particles. Besides that, there are the van der Walls forces between the layers that can induce particle agglomeration, which will negatively affect the desirable properties[8]. Processes to obtain polymer/2D materials nanocomposites with the possibility of scale-up feasibility for industry are extremely important for the manufacture of products with singular properties, keeping low levels of particle content and high levels of dispersion, with the aim to maintain a competitive production cost.

[1] Graphene and Nanomaterials Research Center - MackGraphe, Mackenzie Presbyterian University, Rua da Consolação, 896, São Paulo - SP, 01302-907, Brazil
[2] Instituto de Macromoléculas Professora Eloisa Mano - IMA, Universidade Federal do Rio de Janeiro, Rio de Janeiro - RJ, 21941-970, Brazil
* Correspondent author: guilherminojmf@mackenzie.br (G.J.M. Fechine)

## Strategies to prepare polymer/2D material nanocomposites by melt mixing

The GO and polystyrene were used as templates for the 2D material and polymer, respectively. In order to produce a GO/polymer nanocomposite, two processing methods were tested using a corotational twin screw extruder (Figure 1). The methods are based on solid-solid deposition (SSD) and liquid phase feeding (LPF). The preparation and characterization of graphite, graphite oxide (GrO), and graphene oxide (GO) are presented in the Supplementary Information (SI.01). In SSD, polystyrene (PS) powder (<600 µm) was added to a water dispersion of graphene oxide with few layers (GO) (2 mg.mL$^{-1}$). The mixture was dried in the rotary evaporator equipment, resulting in PS powder covered by exfoliated GO particles. In the case of LPF, a liquid feeder operated by a peristaltic pump with GO water dispersion (1 mg.mL$^{-1}$) was positioned in an extruder at L/D =10, in order to guarantee that the insertion of GO particles occurs with the polymer in a molten state. The pump flow rate was controlled to adjust the GO concentration. In both cases, the focus was to introduce particles of GO already exfoliated into the polymer matrix since the extruder environment is not able to exfoliate graphite oxide to GO.

**A:** Graphite oxide exfoliation in water (2 mg.mL$^{-1}$).

**B:** GO particles are deposited onto PS powder surface using a rotavapor equipment.

**C:** PS powder covered by GO been placed directly into extruder feeder.

**D:** GO suspension is continuously stirred to avoid precipitation at 350 rpm.

**E:** GO suspension is pumped directly onto melted PS at L/D = 10 in a twin-screw extruder.

**F:** Polymeric composites pelletization.

**G:** Specimens production using Injection molding (Type-V ASTM).

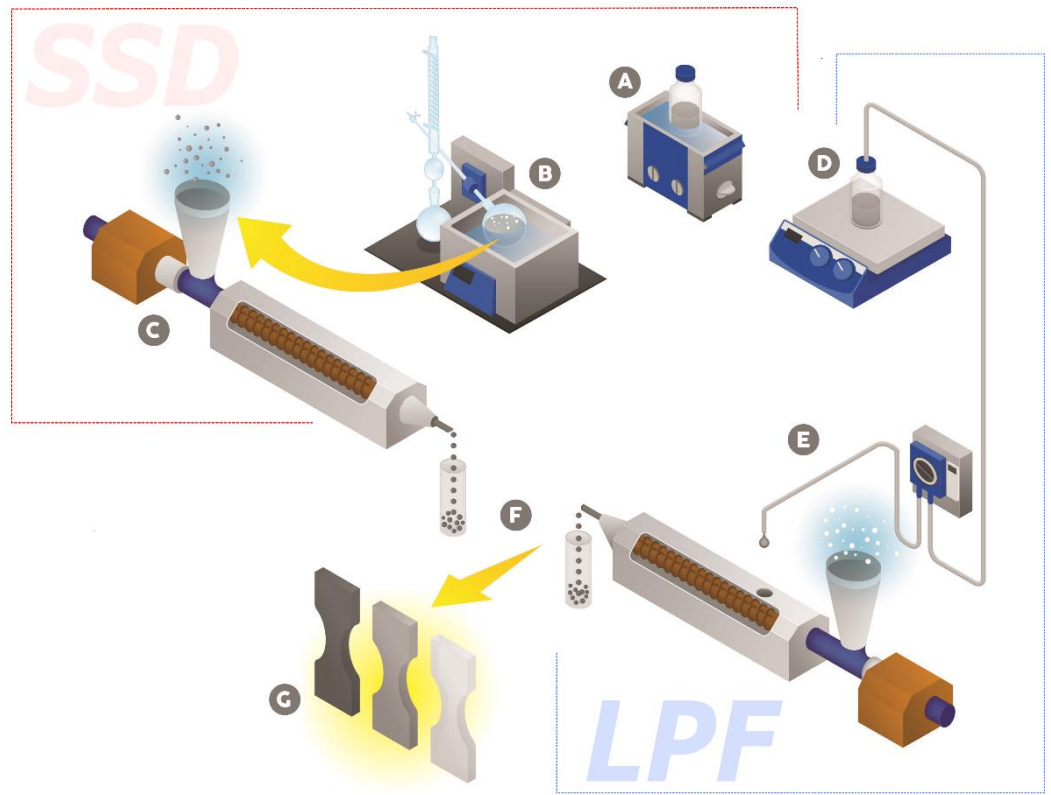

Figure 1. Schematic of SSD and LPF methods, developed for 2D/polymer processing composite at a high production rate.

## Experimental design analysis: Effect of %GO and screw velocity in PS/GO composite mechanical properties

SSD and LPF methods were evaluated using a mixed 2 and 3 levels experimental design in order to understand the effects of screw velocity (SV) and GO particles amount on polymeric mechanical properties. Surface response was obtained using ANOVA with 5% of significant-level and random error distribution. The experimental design and raw data are presented in the SI.02.

According to Figure 2, %GO and SV have different effects on the nanocomposite mechanical performance for both methods. For SSD there is a discontinuous influence of GO content in the Elastic Modulus (E) when a screw velocity of 250 rpm is chosen, in fact, a local minimum at the surface at 0.3% GO is shown. At this point, probably, there is a change in particle aggregation from nano-domains (<0.3% GO) to micro-domains (>0.3% GO), due to particle segregation during processing. However, at high screw velocity level (350 rpm), the E values are higher, and they are not affected by the variation of GO content. This is probably due to better particle distribution and dispersion caused by a higher shear rate generated by the twin-screw operating at this screw velocity. High levels of particle dispersion increase the surface area in contact with the polymer and an interphase zone around each particle is formed modifying the polymer chain mobility[9], and the Young's Modulus is affected positively when compared to processed PS (~4000 MPa). Looking at the influence of screw velocity for all compositions, E tends to increase with higher SV due to better mixing between the particle and the polymeric matrix, which may indicate that lower levels of particle aggregates occur. Note that the data for LPF was completely different, presenting a minimum local of E at low and high screw velocity levels, showing a quadratic dependence of E as a function of GO content.

In contrast to the SSD method, the LPF method presents a decrease of E values for all ranges of GO content due to the increase in screw velocity for all ranges of GO content. According to the method, the GO dispersion is inserted into the extruder at ~240°C generating a very fast evaporation of water molecules, probably, inducing GO particles agglomerates into the system. As the SV increases, the evaporation processes are favored by a mechanical perturbation of the system, and the levels of agglomerates are increased.

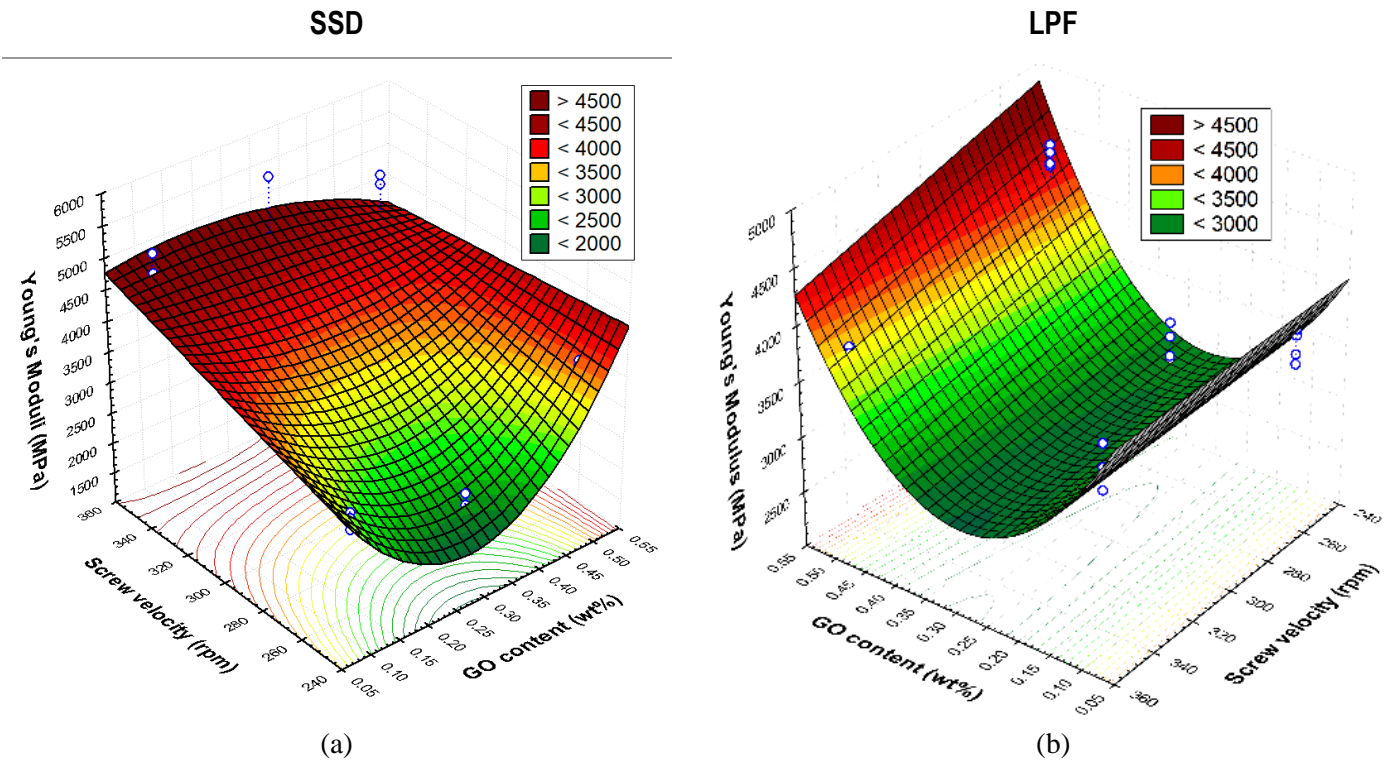

Figure 2. Young's Moduli response surfaces for composites produced by SSD (a) and LPF (b) methods.

Other variables such as UTS (Ultimate Tensile Strength) and SBP (Strain at the Break Point) were also analyzed, however, no good statistical adjustments were obtained (SI.02). The rigidity of PS and defects induced by GO agglomeration must have induced this result. The UTS is strongly influenced by interfacial defects which lead to structure collapses. The same observation is valid for SBP together with the fact that neat PS shows low ductility (SBP < 2%). All data surface is summarized in Table 1. Both methods may be feasible: however, the choice must be made by physicochemical characteristics of the polymer, suitability of the processing variables, and the filler content as well.

A study based on a master-batch (MB) (1.0% of GO) was also performed as an alternative method. The MB was diluted to compositions of 0.1%, 0.3%, and 0.5%. Two MB's were prepared using SSD and LPF methods at 350 rpm screw speed. The MB dilution has followed the experimental design proposed and it was analyzed using ANOVA (SI.03). However, no significant models fit, probably due to the high number of aggregated particles when the MB is prepared and retained even after processing. These aggregates lead to fails (irregularities) on polymer matrix and may cause wide variability in the mechanical response, limiting the use of ANOVA. According to our experimental observation, the production of MB is not a good method to obtain nanocomposites due to increases in GO segregation during processing steps.

Table 1. Regression coefficients of response variables for the different nanocomposite manufacturing routes studied using ANOVA.

| Model: | | \multicolumn{8}{c}{$S = a_0 + a_1(\%GO) + a_2(\%GO)^2 + a_3(SV) + a_4(SV)^2 + a_5(\%GO)(SV) + a_6(\%GO)^2(SV)$} | | | | | | | |
|---|---|---|---|---|---|---|---|---|---|
| Sample | Variable | $a_0$ | $a_1$ | $a_2$ | $a_3$ | $a_4$ | $a_5$ | $a_6$ | $R^2$ |
| | E (MPa) | 3198.7 | -56406.2 | 108187.5 | 4.1 | — | 163.3 | -318.2 | 0.86933 |
| | UTS (MPa) | 27.31640 | -380549 | — | 0.05107 | — | — | -0.04024 | 0.43576 |
| | SBP (%) | 1.1167661 | — | — | 0.000804 | — | — | -0.002649 | 0.35466 |
| | E (MPa) | 4665.2 | -12944.1 | 27895.2 | — | — | — | -15.7 | 0.85028 |
| | TS (MPa) | 60.1861 | — | -15.0744 | -0.0500 | — | — | — | 0.75152 |
| | SBP (%) | 2.794921 | — | — | -0.004351 | — | -0.001745 | — | 0.52037 |

**Processing effects on composites structure**

Once different mechanical responses were obtained for the composites, a study regarding structural changes was performed in order to relate processing, structure, and properties. The first structural change studied was the PS molecular weight change during the extrusion process. All raw molecular weight data is presented in the SI.04. The mechanical response of polymeric materials is length chain dependent, once it influences chain entanglement changing properties in elastic and plastic deformation regions. Molecular weight data of neat PS (Figure 2a) shows a slight reduction in $\bar{M}n$ and $\bar{M}w$ after extrusion caused by high shear rate and thermo-oxidation reactions that occurred duringprocessing[10–12]. Values of $\bar{M}n$ and $\bar{M}w$ of PS show a reduction for all nanocomposites prepared via the SSD methodology at 350 rpm when compared with neat PS, but they are higher than processed PS (Figure 2b). The stabilizer effect may be a result from a GO lubricant effect during processing. Figure 2c shows the difference in $\bar{M}w$ compared to processed PS for different routes with 0.5% of GO and 350 rpm of SV. Again, the effect of less reduction of $\bar{M}w$ for all nanocomposites is observed.

For nanocomposites obtained by master-batch dilution (MB), the level of polymeric chain preservation was lower than materials directly obtained (SSD and LPF), due to the higher number of processing steps. The behavior of all materials at 250 rpm of SV is quite similar (SI.04).

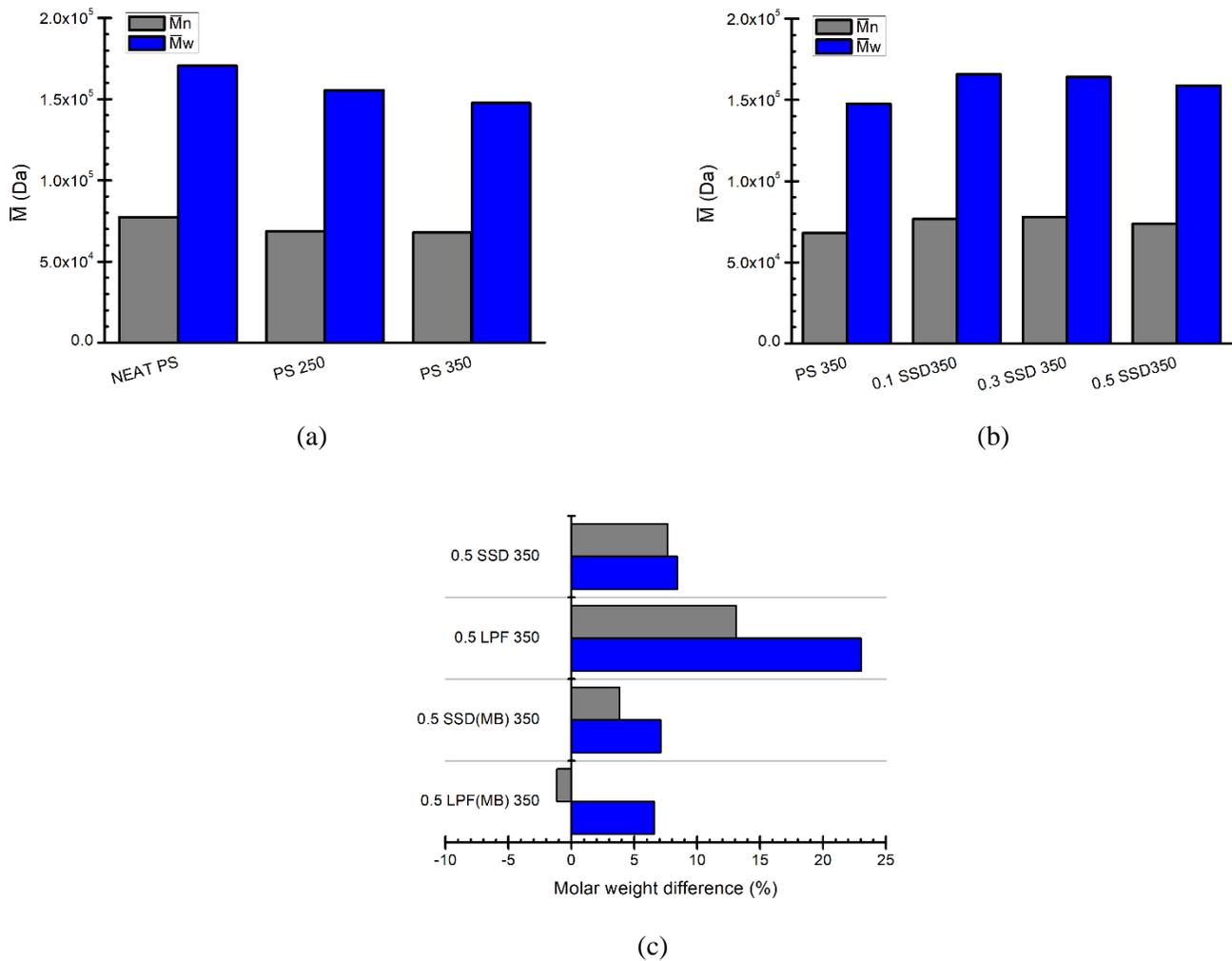

Figure 2. (a) Molecular weight - $\overline{M}n$ and $\overline{M}w$ of neat PS and processed PS. (b) Effect of %GO in $\overline{M}n$ and $\overline{M}w$ of PS processed using SSD at 350 rpm. (c) $\overline{M}w$ difference to all routes at 0.5% GO compared to processed PS. Sample code: XX YY ZZZ (XX = GO content, YYY = method and ZZZ = screw velocity).

Particle dispersion and distribution onto the polymer matrix were analyzed using X-ray microtomography (XR-MT) and low-field nuclear magnetic resonance (LF-NMR). These techniques are beyond micro and nanoscale, giving the possibility to evaluate the morphology for millimeter samples size. All XRMT and LF-NMR data is presented in the SI.04. The object density (OD) of the particles was obtained from XR-MT, and its increase means the tendency of particle agglomeration. There is a reduction tendency of spin-lattice relaxation time (with a time constant - $T_1H$) with an increase in object density (OD) taking out MB dilution materials (Figure 3a). A high level of particle aggregation means less surface contact between the polymer and GO, which consequently means the polymer chains are free to move. Three images generated from XR-MT are presented in Figure 3a confirming that high SV leads to a very low level of aggregates, while samples prepared by master-batch possess an enormous volume of them. According to the data shown in Figure 3b, it is possible to affirm that high E values are compatible with high $T_1H$, indicating a better interfacial adhesion between the polymer matrix and GO particles. The polymer chains anchor onto GO particles and the spin-lattice relaxation time is increased. An increase in $T_1H$ values with SV, as well as an increase in elastic modulus (as shown before) for composites prepared at an SV of 350 rpm, indicate a better GO/polymer interface and a low level of particle aggregation (low number of OD, Figure 3a). Materials prepared using MB dilution do not show any correlation between OD and $T_1H$. The great quantity and volume of agglomerates turn the composite into a material with a heterogeneous structure. Elastic Modulus data does not show any correlation with $T_1H$ values, also indicating a non-homogeneous structure of materials prepared using MB dilution. There is no direct relation between observed $T_1H$ values and molecular weight (SI.04), and when plotting UTS and SBP as $T_1H$ values function no correlation was found. As mentioned before, the brittleness of PS and structural fails caused by agglomerates formed during processing have a great effect on a mechanism involving plastic deformation due to lack of matrix-particles interaction.

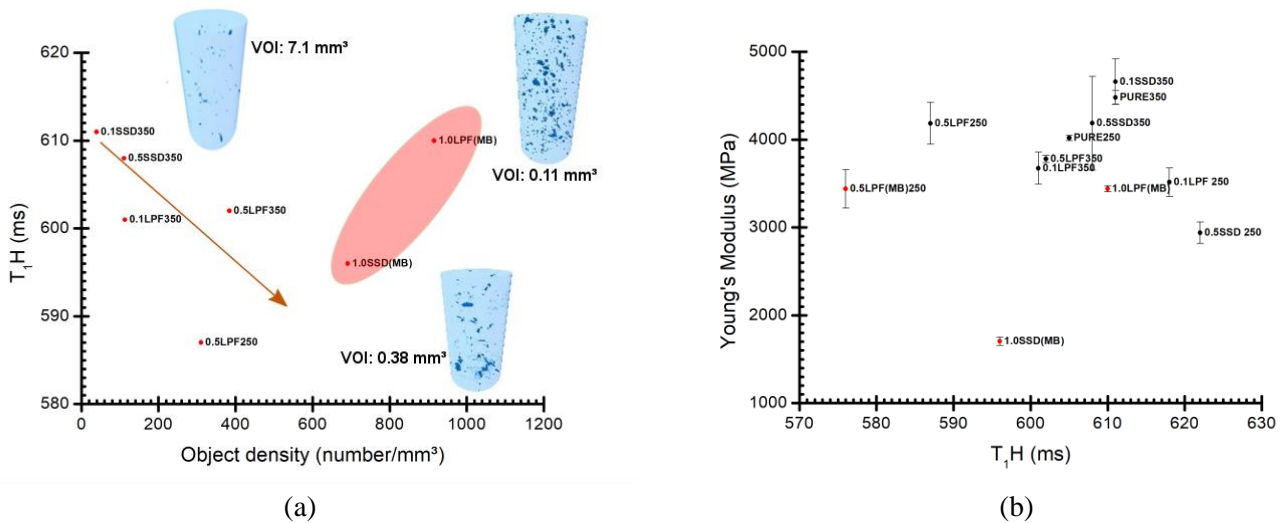

(a) (b)

Figure 3. Spin-lattice relaxation time ($T_1H$) correlations with: (a) density of aggregates formed into composite and (b) Elastic Modulus. VOI: Volume of interest. Sample code: XX YY ZZZ (XX = GO content, YYY = method and ZZZ = screw velocity).

In order to obtain indications regarding exfoliation and dispersion levels of the particles as a function of the processing parameters, rheological tests in steady shear flow were performed. Figure 4 presents the steady shear viscosity versus the shear rate for neat PS and its composites at the two different screw velocities used to process the composites by SSD method. In Figure 4a, the viscosity behavior of the materials processed at 250 rpm of SV is quite similar presenting a shear-thinning behavior (i.e., pseudoplastic behavior), with the shear viscosity decreasing with the shear rate. This behavior is due to the disentanglement of the molecular chains and 2D particles alignment/sliding under high shear rate[13]. Furthermore, it should be considered that each polymeric chain can present different lengths, and consequently, the steady shear flow develops alignment/disentanglements of the chains with the increase in the shear rate. Therefore, a Newtonian region in the experimental shear rate window was not detected. Polymeric chains and 2D particle arrangements can change during flow process, since entanglements and 2D particle dispersion degrees/sizes are not homogeneous in the melt state, as indicated by the variations in viscosity slope. Note that in Figure 4a, above 0.1 s$^{-1}$ the shear viscosity for nanocomposites tends to be smaller than pure polymer, which can suggest a lubrication effect provided by the 2D particle[14,15]. The graphene oxide dispersed in the polymeric matrix can display some stacked layers, and the lubrication mechanism can be based on the assumption that the GO layers are able to slide over each other with the increase of shear flow, being associated to the cards pack[16–18]. This mechanism is detected in the nanocomposites obtained at lower screw velocity (250rpm) suggesting a lower exfoliation degree and occurred at a lower shear rate for larger concentrations. This observation indicates that the lubrication effect is less evident in the presence of exfoliated particles (nanometric scale) as it is seen in Figure 4b (screw velocity of 350rpm). In this case, there is the intimate contact between the polymeric chains and the exfoliated GO layers, and consequently, it can induce a modification in the polymeric relaxation dynamics[19,20]. This behavior can be attributed to the formation of a confinement structure of exfoliated GO layers, hindering the disentangling of the polymeric chain during flow, i.e., the exfoliated layers prevent a free-rotation of molecular chains and dissipation of stress, inducing the increase of shear viscosity[21,22] as observed in Figure 4b. Note that the level of agglomerates tends to affect the rheological response, and consequently, affect the final properties of the polymeric nanocomposite. It can be seen that the screw velocity will influence the dispersion and level of agglomerates. The transmission electronic microscopy image of the PS nanocomposites with a concentration of 0.1 wt% and processing at 350 rpm screw velocity displayed a distribution of thin graphene oxide layers, as shown in Figure 4c. It can also be observed that there are some regions in which GO flakes are folded, as indicated by arrows. In the case of the PS/GO with a concentration of 0.3 wt% (see Figure 4d), the GO layers are parallel to the main plane of the polymeric nanocomposite film, although a few layers are oriented in the other direction. It should also be considered that there are flakes in the nanoscale, that are not able to be seen, suggesting the high exfoliated level of GO. It is recognized that the extruder process parameters are important variables that must be optimized to obtain a high dispersion degree and a desirable final level of agglomerates, with the aim to obtain the final property designed. In general, a great effect of screw velocity that leads to a smaller volume of the 2D particles agglomerates, as shown by microtomography was observed. These experimental observations lead to the conclusion that nanocomposites formed by a polymer/2D particle with nanometric scale, specifically GO, are just viable if the particle concentration is low, to avoid aggregates in micrometric scale. The strategies used here for graphene oxide and polystyrene may be also used to manufacture nanocomposites based on other 2D materials and several polymers. Others polymer/2D material composites systems obtained by using the strategies proposed here are presented in the SI.05. Mechanical properties (tensile test) were used to evaluate the feasibility of composite production methods.

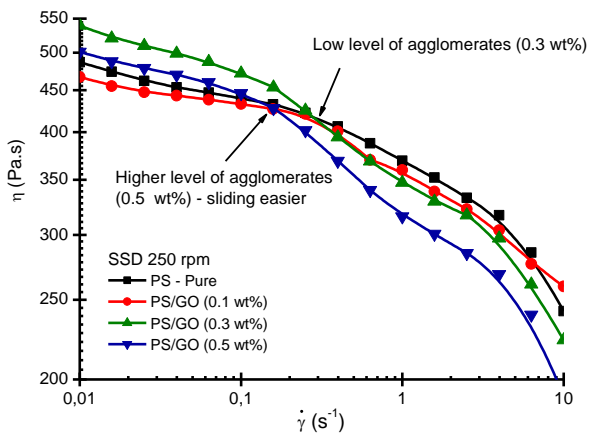

(a)

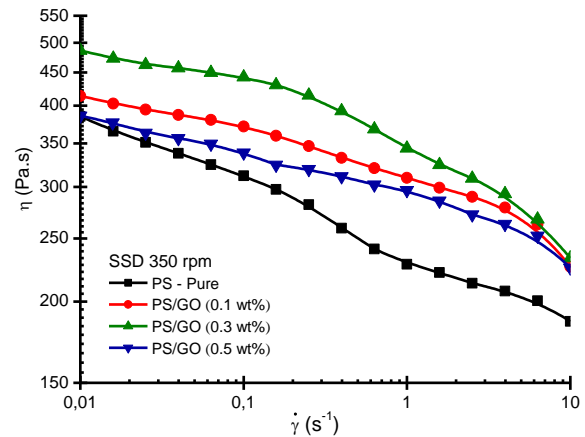

(b)

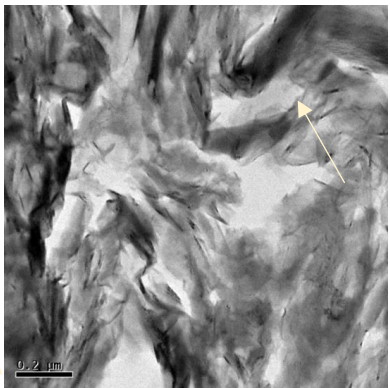

(c)

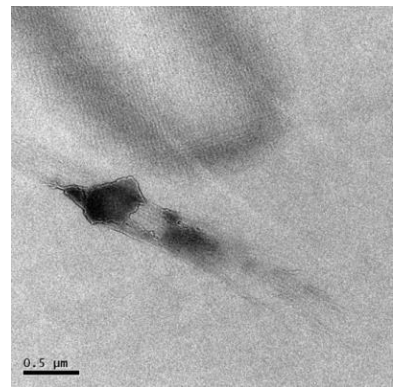

(d)

Figure 4 – Shear rate dependence of the steady shear viscosity at 230°C for polymer pure and its nanocomposites (0.1; 0.3 and 0.5 %wt): a) SSD at 250 rpm screw speed; b) SSD at 350 rpm screw speed. TEM micrographs of the nanocomposites obtained by SSD at 350 rpm screw speed: c) PS with 0.1 wt% GO and d) PS with 0.3 wt% GO.

## Conclusion

Here, for the first time, a complete study showing how a 2D material (graphene oxide) is inserted into the polymeric matrix in order to produce a nanocomposite by using a conventional polymer processing machines leading to large-scale manufacturing is presented. A better understanding of morphology was reached since traditional analysis (mechanical test, molecular weight, rheology, and transmission electron microscopy) and unusual techniques for this type of study (XR-MT and F-NMR) were used. According to the results presented here, it is possible to obtain 2D based polymer nanocomposites using pre-exfoliated particles added directly into an extruder (LPF) or deposited on the polymer powder surface (SSD) before extrusion. Using pre-exfoliated 2D material is key to obtaining a good dispersion of particles into the polymer matrix. Both methodologies were used to prepare master-batches, however, this conventional approach is not indicated for 2D based polymer nanocomposites. This is because of the high level of particle aggregation, which occurs during master-batch production even for 1.0% w/w of content. The reason for that is the high surface area of these particles.

The methodologies described here are extendable to all polymers that can be processed using melt mixing. We demonstrate that polymers and 2D materials can be mixed using standard and industrially scalable polymer processing, providing nanocomposites with suitable morphology to obtain desirable properties.

## Method

*Preparation of GO suspension.* Firstly, the graphene oxide was prepared following Hummer´s method[23]. 200 mg of graphite oxide (GrO) was exfoliated into deionized water, using an ultrasonic bath (Elma, P30) for 30 min in individual 100 mL batches, in order to obtain a 2 mg.mL$^{-1}$ GO suspension. To avoid precipitation, the GO suspension was constantly stirred (~350 rpm) until the required suspension was obtained.

*Solid-Solid Deposition (SSD).* For this method, the GO suspension was dried together with PS powder into a rotavapor equipment at 55° C. The amount of GO suspension used was compatible with the final desired GO content in the composite (0.1%, 0.3%, 0.5%, and 1%).

*Liquid Phase Feeding (LPF).* For this method, the GO suspension was directly added into the extruder using an opening at L/D = 10 (to guarantee that polymer is molten). The GO suspension feeding rate was set to obtain the desired GO content in the composite (0.1%, 0.3%, 0.5% and 1%), using a peristaltic pump.

*Extrusion parameters.* Both methods were designed to produce GO composites through extrusion, because it is one of the most used processing techniques. All samples were extruded in a co-rotating twin screw extruder (Process 11, ThermoScientific) with L/D = 40 and D = 11 mm. A 4 g.min$^{-1}$ feeding rate was adopted for all runs and the temperature profile used was: 170/230/250/250/260/230 °C from hopper to die.

*Injection molding.* Test specimens were produced using a barrel mini-inject (MiniJet Pro, ThermoScientific) using an ASTM Type V test specimen mold. Temperatures of 230° C and 60 °C, for barrel and mold, respectively. The injection pressure of 200 MPa (30 s) and post-pressure of 150 MPa (20 s) were applied.

*Mechanical properties measurement.* Were performed in a universal test machine Zickwick Machine at deformation rate of 1 mm.min$^{-1}$ at room temperature (25° C).

*Molecular weight measurements.* Molecular weight and its distribution were measured using a size exclusion chromatographer (HT-GPC 305A, Malevern) equipped with an RI detector. All samples were analyzed in THF at 40° C with 0.25% of BHT. A 1 mL.min$^{-1}$ solvent flow was used in both pumps and calibration was performed using mixed PS standards (PolyAnalytik) from 1.5 kDa to 4 MDa. All samples were filtered using 0.45 µm syringe filter.

*Atomic Force Microscopy (AFM).* Drop casting of GO dispersion was prepared on top of fresh mica and analyzed in an Icon Dimension (Bruker) equipped with RTESPA. In order to obtain a good statistic, more than 800 particles were counted and measured using Gwyddion Software.

*Raman confocal microscopy.* Raman spectra were acquired with a WITec Alpha 300R confocal Raman spectrometer. The excitation source was a 532 nm laser.

Thermo-Gravimetric Analysis (TGA). The thermal stability of graphite and graphite oxide was characterized using thermogravimetric analysis (DSC/TGA Q600, TA Instruments). All measurements were conducted under a dynamic nitrogen flow over a temperature range of 30–1000 ºC with a slow ramp rate of 10 ºC min.

*X-ray Microtomography.* Pieces of at least 2mm x 2mm x 2mm from tensile strength specimens were used for the characterization. Samples were analyzed in a SkyScanner

1272 (Bruker), using 20kV and 175 µA X-ray source, with a final image resolution of 2 µm/pixel.

*Low-Field Nuclear Ressonance.* The proton spin-lattice relaxation time $T_1H$ of the sample was analyzed in a Maran Ultra 23 (Oxford), operating at 23 MHz for protons, equipped with 18mm NMR tube. The pulse sequence used to determine the relaxation data was inversion-recovery. The 90 degree was 7.6µs, which was automatically calibrated by the equipment software. The amplitude of the FID was sampled for 40 τ data point varying from 0.1 to 10000 ms with four scans each and 10s of recycle delay. The values of $T_1H$ were obtained by fitting the exponential data using the Winfit program that comes with the spectrometer.

*Rheological test by steady state.* Discoidal specimens were prepared using a press mold at 230°C, with 3 tons of pressure for 2 minutes. Rheological tests in rotational flow were performed at Anton Paar 102 rheometer with controlled strain, using plate/plate geometry, 1 mm gap at 230° C and shear rate from $0.01$ $s^{-1}$ to $10$ $s^{-1}$.

*Transmission Electron Microscopy (TEM).* TEM imaging was performed on a JEOL 1200 EXII microscope at 80 keV. Samples with 60 nm thickness were collected on top of 200 mesh-copper grids.

**References**
1. Novoselov, K. S. *et al.* Electric field effect in atomically thin carbon films. *Science* **306,** 666–9 (2004).
2. Kim, H., Abdala, A. A. & Macosko, C. W. Graphene/polymer nanocomposites. *Macromolecules* **43,** 6515–6530 (2010).
3. Potts, J. R., Dreyer, D. R., Bielawski, C. W. & Ruoff, R. S. Graphene-based polymer nanocomposites. *Polymer (Guildf).* **52,** 5–25 (2011).
4. Zhao, X., Zhang, Q., Chen, D. & Lu, P. Enhanced mechanical properties of graphene-based polyvinyl alcohol composites. *Macromolecules* **43,** 2357–2363 (2010).
5. Yousefi, N. *et al.* Highly aligned, ultralarge-size reduced graphene oxide/polyurethane nanocomposites: Mechanical properties and moisture permeability. *Compos. Part A Appl. Sci. Manuf.* **49,** 42–50 (2013).
6. Kim, H., Miura, Y. & Macosko, C. W. Graphene/polyurethane nanocomposites for improved gas barrier and electrical conductivity. *Chem. Mater.* **22,** 3441–3450 (2010).
7. P. Pokharel, D. L. Thermal and Mechanical Properties of Reduced Graphene Oxide/Polyurethane Nanocomposite. *J. Nanosci. Nanotechnol.* **14,** 5718–5721 (2014).
8. Novoselov, K. S. *et al.* Two-dimensional atomic crystals. *Proc. Natl. Acad. Sci. U. S. A.* **102,** 10451–10453 (2005).
9. RamanathanT. *et al.* Functionalized graphene sheets for polymer nanocomposites. *Nat. Nanotechnol.* **3,** 327–331 (2008).
10. Oliveira, C. F. P., Carastan, D. J., Demarquette, N. R. & Fechine, G. J. M. Photooxidative behavior of polystyrene-montmorillonite nanocomposites. *Polym. Eng. Sci.* **48,** (2008).
11. Fechine, G. J. M., Rabello, M. S. & Souto-maior, R. M. The effect of ultraviolet stabilizers on the photodegradation of poly ( ethylene terephthalate ). *Polym. Degrad. Stab.* **75,** 153–159 (2002).
12. Al-Itry, R., Lamnawar, K. & Maazouz, A. Improvement of thermal stability, rheological and mechanical properties of PLA, PBAT and their blends by reactive extrusion with functionalized epoxy. *Polym. Degrad. Stab.* **97,** 1898–1914 (2012).
13. Chen, D., Yang, H., He, P. & Zhang, W. Rheological and extrusion behavior of intercalated high-impact polystyrene/organomontmorillonite nanocomposites. *Compos. Sci. Technol.* **65,** 1593–1600 (2005).
14. Rapoport, L. *et al.* Polymer Nanocomposites with Fullerene-like Solid Lubricant. *Adv. Eng. Mater.* **6,** 44–48 (2004).
15. Pu, J. *et al.* Preparation and tribological study of functionalized graphene-IL nanocomposite ultrathin lubrication films on si substrates. *J. Phys. Chem. C* **115,** 13275–13284 (2011).
16. Wang, S., Chen, Y., Ma, Y., Wang, Z. & Zhang, J. Size effect on interlayer shear between graphene sheets. *J. Appl. Phys.* **122,** 74301 (2017).
17. Areshkin, D. A. & White, C. T. Building blocks for integrated graphene circuits. *Nano Lett.* **7,** 3253–3259 (2007).
18. Dorri Moghadam, A., Omrani, E., Menezes, P. L. & Rohatgi, P. K. Mechanical and tribological properties of self-lubricating metal matrix nanocomposites reinforced by carbon nanotubes (CNTs) and graphene - A review. *Compos. Part B Eng.* **77,** 402–420 (2015).
19. Magomedov, G. M., Khashirova, S. Y., Ramazanov, F. K., Beslaneeva, Z. L. & Mikitaev, a. K. Relaxation properties and structures of polymer nanocomposites based on modified organoclays. *Polym. Sci. Ser. A* **56,** 652–661 (2014).
20. Comer, A. C., Heilman, A. L. & Kalika, D. S. Dynamic relaxation characteristics of polymer

nanocomposites based on poly(ether imide) and poly(methyl methacrylate). *Polymer (Guildf).* **51,** 5245–5254 (2010).
21. Krishnamoorti, R., Vaia, R. A. & Giannelis, E. P. Structure and dynamics of of polymer-layered silicate nanocomposites. *Chem Mater* **8,** 1728–1734 (1996).
22. Lim, Y. T. & Park, O. O. Rheological evidence for the microstructure of intercalated polymer/layered silicate nanocomposites. *Macromol. Rapid Commun.* **21,** 231–235 (2000).
23. Stankovich, S. *et al.* Synthesis of graphene-based nanosheets via chemical reduction of exfoliated graphite oxide. *Carbon N. Y.* **45,** 1558–1565 (2007).

# Supplementary Information

## SI.01 - Graphene Oxide preparation and characterization

Graphene oxide (GO) was obtained from the exfoliation of graphite oxide (Gr-O) in water. The Gr-O was prepared by oxidizing the graphite flakes (Gr) using a modified Hummer's method. The materials were characterized using X-ray diffraction (XRD), thermogravimetric analysis (TGA), and Raman spectroscopy, as shown in Fig. S1. The first indication of success in Gr-O synthesis is the change in the XRD pattern observed in Fig.S1a, where Gr-O presents a peak in 11.34°, caused by an increase in distance of (002) planes of the graphitic structure, due to the oxygenated groups formed during oxidation along the structure. Using TGA (Fig.S1b), it is possible to observe a significant difference between Gr and Gr-O. The weight loss in the region between 200° C and 450° C for Gr-O is the result of the lost oxygenated groups present in the Gr-O structure, while the Gr weight loss is only observed above 700° C. Raman spectroscopy (Fig.S1c) was used in order to evaluate characteristics of graphite material, mainly D band. It is possible to observe that Gr-O presents an increase in D peak (1352.14 cm$^{-1}$), which is related to the breaking of symmetry in graphite when it is oxidized.

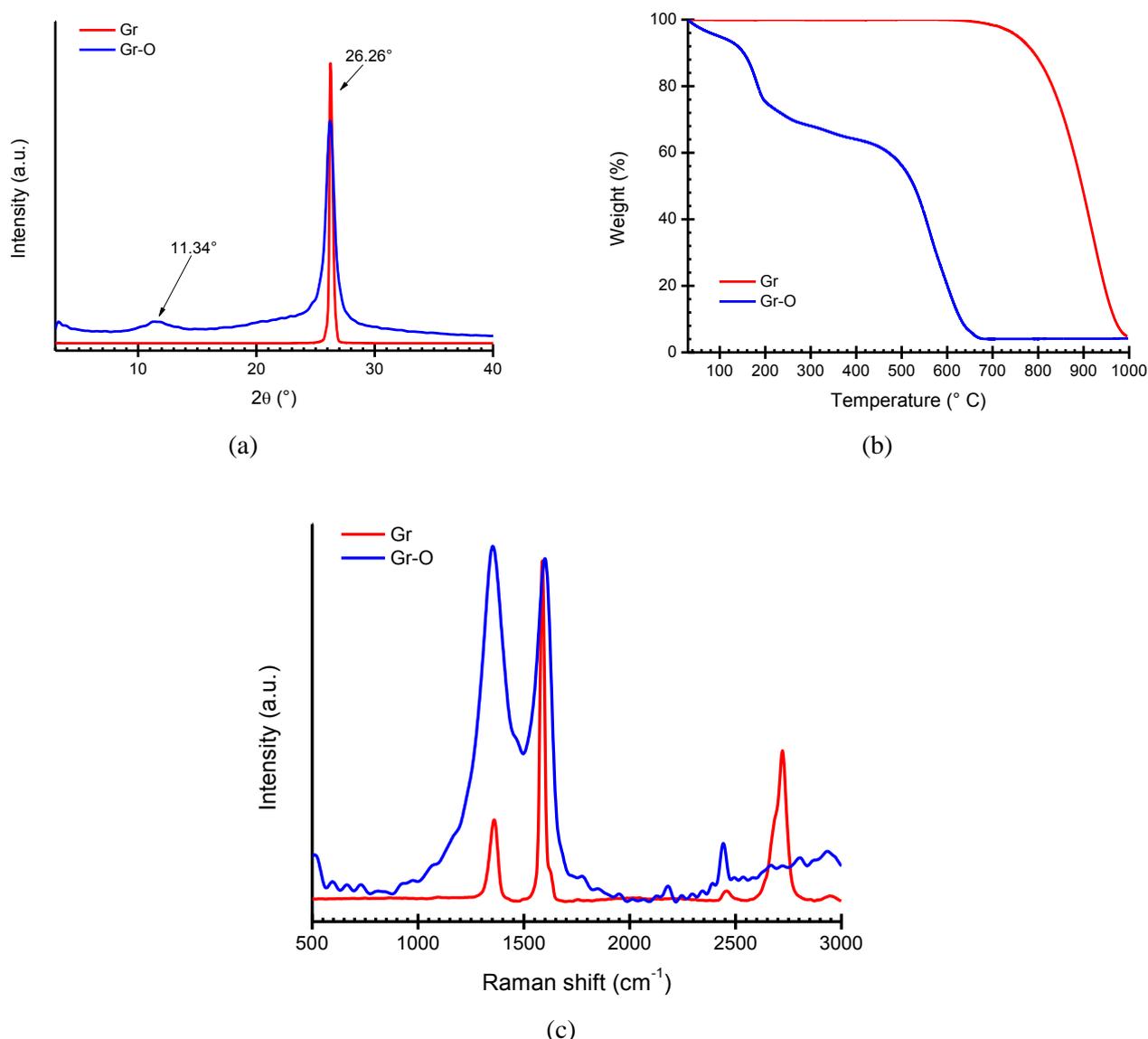

**FIGURE.S1** – Gr-O characterizations. (a) XRD pattern. (b) Weight loss in an inert atmosphere (heating rate: 10° C/min). (c) Raman spectra (excitation laser : 532 nm).

Once graphite oxidation has been carried out successfully, graphene oxide was obtained from the exfoliation of Gr-O in water (1 mg/mL, 100 mL batches) in a sonication bath for 30 minutes (300 W). In order to verify the flakes morphology, aliquots of GO suspension were dropped onto a fresh mica surface and then analyzed using AFM (Fig. S2). Through AFM measurements, the thickness and lateral size of GO particles were obtained, using Gwyddion Software. Fig S2a shows one of the analyzed areas, and Fig.S2b shows painted particles performed in Gwyddion.

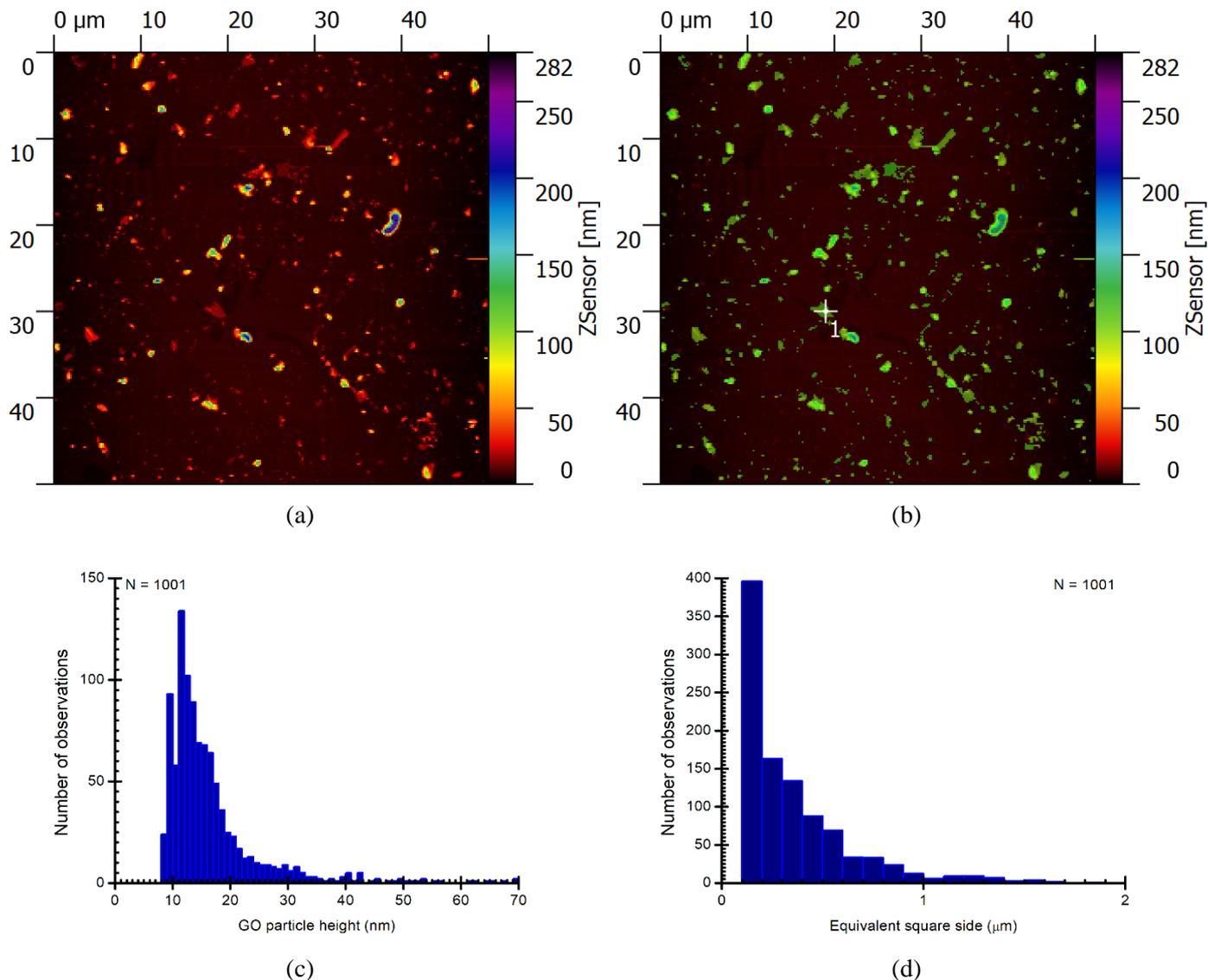

**FIGURE.S2** — GO suspension characterization. (a) AFM z-sensor mapping (50 µm x 50 µm). (b) Grain delimitation (green area) performed using Gwyddion. (c) Particle height distribution. (d) Lateral size estimation distribution (equivalent square side).

GO AFM analysis shows that few-layer-GO (around 10 to 40 layers, assuming 0.75nm for the distance between layers) were obtained after Gr-O exfoliation (Fig.S2c), indicating a very good Gr-O exfoliation. At the same time, an import reduction in the lateral size of GO particles was observed, showing a predominant presence of particles lower than 1 µm. The size of graphite particles used here was about 150 µm.

### SI.02 - 2D-basead nanocomposites production

The study of methods to produce 2D-based polymer nanocomposite was performed using a mixed 2-3 level experimental design assisted by ANOVA. The effects of GO content (%GO) and screw velocity (SV) were evaluated as shown in Tab.S1.

**TABLE.S1** — Processing PS/GO composite experimental design.

| Run | % GO | SV | % GO (wt%) | SV (rpm) |
|---|---|---|---|---|
| 1 | + | + | 0.5 | 350 |
| 2 | + | - | 0.5 | 250 |
| 3 | 0 | + | 0.3 | 350 |
| 4 | 0 | - | 0.3 | 250 |
| 5 | - | + | 0.1 | 350 |
| 6 | - | - | 0.1 | 250 |

In Fig.S3a, it is possible to observe that there is no negative impact of screw velocity on Young's Modulus when the PS is processed. Analyzing composites processed at 250 rpm (Fig.S3b), there is no sample with significantly high values of Young's Modulus when compared to the processed PS. On the other hand, composites processed at 350 rpm showed a Young's Moduli higher than PS processed when the SSD approach was used. The effect of an increase in Young's Moduli is caused by the presence of GO with a significantly dispersed particle distribution due to an increase in shear rate (screw velocity) that also leads to a better interface development.

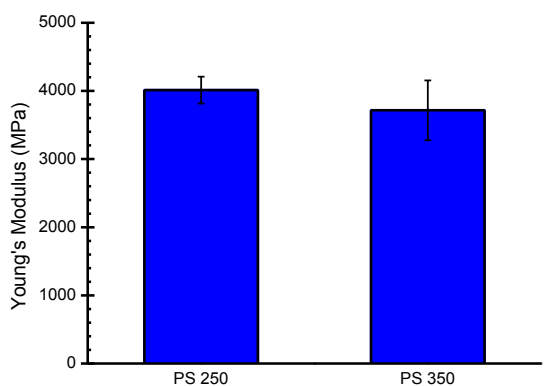

(a)

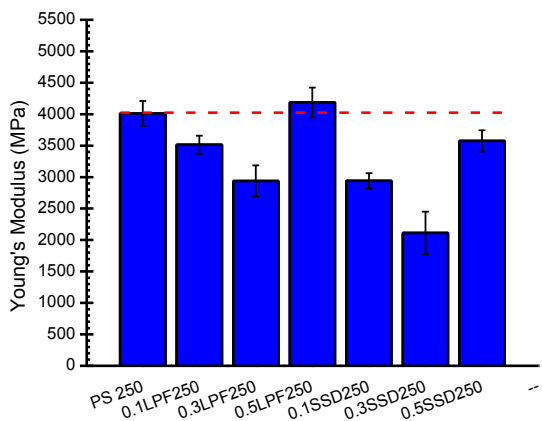

(b)

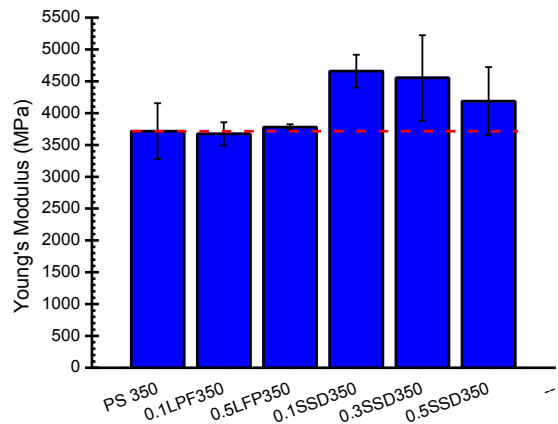

(c)

**FIGURE.S3** — (a) Effect of screw velocity on Young's Modulus for processed PS. (b) Comparison of Young's Modulus between composites and PS both processed at 250 rpm. (c) Comparison of Young's Modulus between composites and PS both processed at 350 rpm. Sample code: XX YY ZZZ (XX = GO content, YYY = method and ZZZ = screw velocity).

The UTS (Ultimate Tensile Strength) and SBP (Strain at the Break Point) data shows high variations, and the $R^2$ values are quite low for these two mechanical properties (Tab.S2 — Tab.S5). This high variation is probably due to GO agglomerates obtained during processing. Polystyrene is a brittle polymer, and the presence of agglomerates increases the probability of a break caused by interfacial defects. Residual plots (Fig.S3) show that only Young's Moduli presents a random error distribution while UTS and SBP present a quasi-linear increase in residuals with the increase in observed values, i.e., the model error increases in a non-random way creating an error tendency in the statistical model. Once statistical models for UTS and SBP did not present random error distribution, this surface could not be considered valid. On the other hand, Young's Moduli models are quite good, and the effects of variables on processing are summarized in Fig.S5 with Pareto charts. According to the Pareto charts, SSD and LPF present different responses for the variables studied. The SSD method is strongly affected by SV, probably because the shear rate generated into extrudes helps to exfoliate the deposited GO.

**TABLE.S2** — Mean and standard deviation in TS and SBP for PS/GO composites produced via SSD.

| Run | % GO (wt%) | SV (rpm) | UTS (MPa) | SBP (%) |
|---|---|---|---|---|
| 1 | 0.1 | 250 | 38.6 ± 4.3 | 1.27 ± 0.05 |
| 2 | 0.3 | 250 | 41.7 ± 3.4 | 1.38 ± 0.05 |
| 3 | 0.5 | 250 | 34.6 ± 3.3 | 1.17 ± 0.15 |
| 4 | 0.1 | 350 | 45.2 ± 2.8 | 1.46 ± 0.23 |
| 5 | 0.3 | 350 | 41.1 ± 2.8 | 1.24 ± 0.05 |
| 6 | 0.5 | 350 | 40.9 ± 3.2 | 1.22 ± 0.11 |
| | $R^2$ | UTS (MPa) | 0.4358 | |
| | $R^2$ | SBP (%) | 0.2608 | |

**TABLE.S3** — Mean and standard deviation in TS and SBP for PS/GO composites produced via LPF.

| Run | % GO (wt%) | SV (rpm) | UTS (MPa) | SBP (%) |
|---|---|---|---|---|
| 1 | 0.1 | 250 | 49.8 ± 0.7 | 1.74 ± 0.17 |
| 2 | 0.3 | 250 | 43.9 ± 2.0 | 1.67 ± 0.39 |
| 3 | 0.5 | 250 | 44.4 ± 2.0 | 1.36 ± 0.11 |
| 4 | 0.1 | 350 | 42.3 ± 1.4 | 1.13 ± 0.14 |
| 5 | 0.3 | 350 | — | — |
| 6 | 0.5 | 350 | 39.7 ± 1.6 | 1.20 ± 0. |
| | $R^2$ | UTS (MPa) | 0.5619 | |
| | $R^2$ | SBP (%) | 0.4488 | |

**TABLE.S4** — Mean and standard deviation in TS and SBP for PS/GO composites produced via SSD-MB.

| Run | % GO (wt%) | SV (rpm) | UTS (MPa) | SBP (%) |
|---|---|---|---|---|
| 1 | 0.1 | 250 | 44.1 ± 4.6 | 1.58 ± 0.23 |
| 2 | 0.3 | 250 | 43.1 ± 2.5 | 1.34 ± 0.11 |
| 3 | 0.5 | 250 | 43.1 ± 2.5 | 1.34 ± 0.11 |
| 4 | 0.1 | 350 | 43.9 ± 3.0 | 1.43 ± 0.14 |
| 5 | 0.3 | 350 | 47.1 ± 0.8 | 1.68 ± 0.66 |
| 6 | 0.5 | 350 | 40.0 ± 10.4 | 1.29 ± 0.48 |
| | $R^2$ | UTS (MPa) | 0.0401 | |
| | $R^2$ | SBP (%) | 0.0513 | |

**TABLE.S5** — Mean and standard deviation in TS and SBP for PS/GO composites produced via LPF-MB.

| Run | % GO (wt%) | SV (rpm) | UTS (MPa) | SBP (%) |
|---|---|---|---|---|
| 1 | 0.1 | 250 | 41.2 ± 4.5 | 1.29 ± 0.30 |
| 2 | 0.3 | 250 | 40.3 ± 2.3 | 1.50 ± 0.19 |
| 3 | 0.5 | 250 | 40.8 ± 2.7 | 1.26 ± 0.17 |
| 4 | 0.1 | 350 | 38.1 ± 2.0 | 1.03 ± 0.16 |
| 5 | 0.3 | 350 | 42.2 ± 0.8 | 1.34 ± 0.13 |
| 6 | 0.5 | 350 | 40.2 ± 3.5 | 1.25 ± 0.06 |
| | $R^2$ | UTS (MPa) | 0.0067 | |
| | $R^2$ | SBP (%) | 0.0075 | |

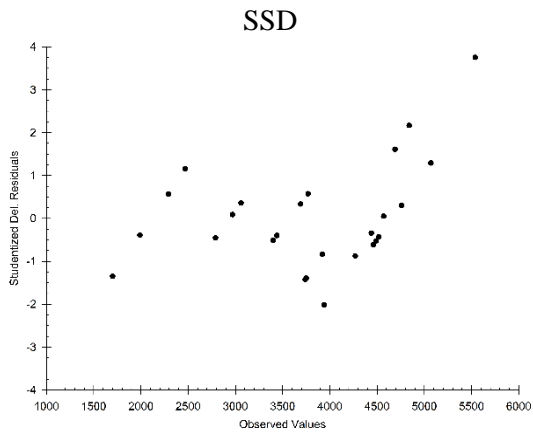
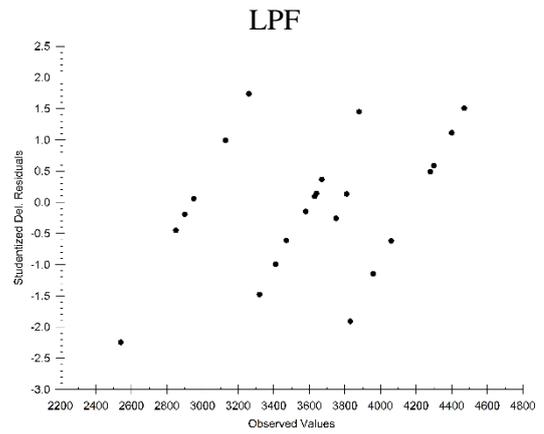
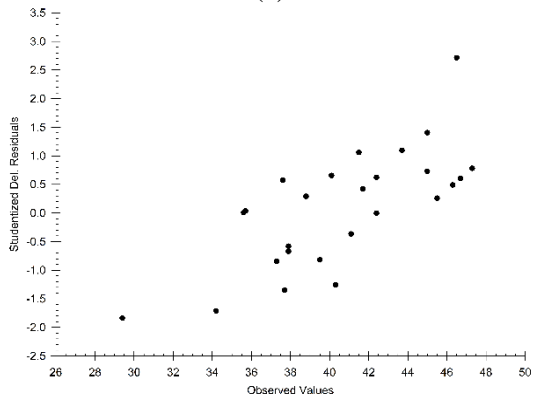
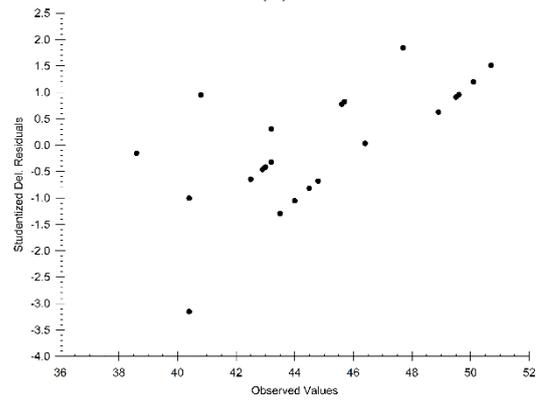
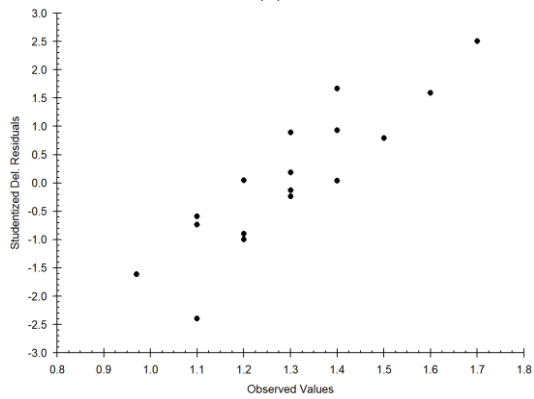
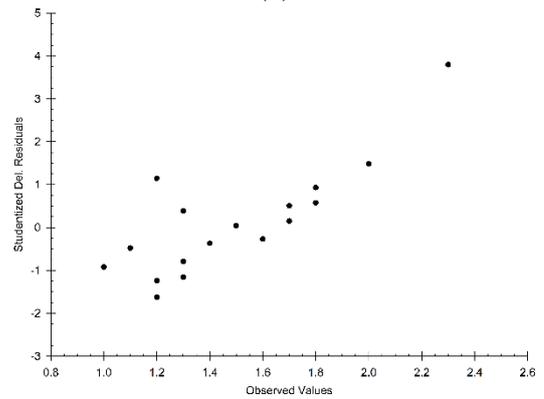

**FIGURE.S4** — Residual plots of SSD surface responses for: (a) Young's Moduli, (c) UTS, and (e) SBP. Residual plots of LPF surface responses for: (b) Young's Moduli, (d) UTS, and (f) SBP.

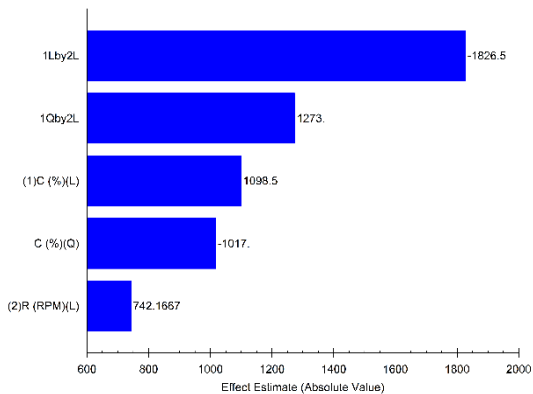
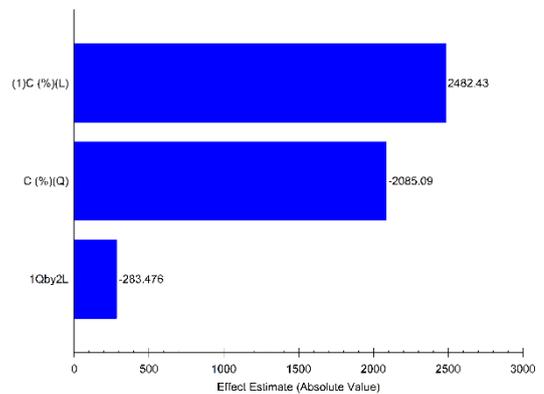

**FIGURE.S5** — Pareto effect chart for Young's Moduli: (a) SSD and (b) LPF processing. 1: GO content (%GO), 2: Screw Velocity; L: linear effect and Q: Quadratic effect.

## SI.03 - Master-batch dilution

The approach using a master-batch (MB) dilution (1.0% GO) from SSD and LPF, obtained at 350 rpm was also evaluated. All runs were based on the experimental design and evaluated using ANOVA analyzing GO content and screw velocity effects. As can be seen in Fig.S6, for both methods there is no good fit to regression models. This probably occurred due to the high volume of GO aggregates that are formed during processing. This hypothesis is supported by X-ray tomography data (Tab.S6).

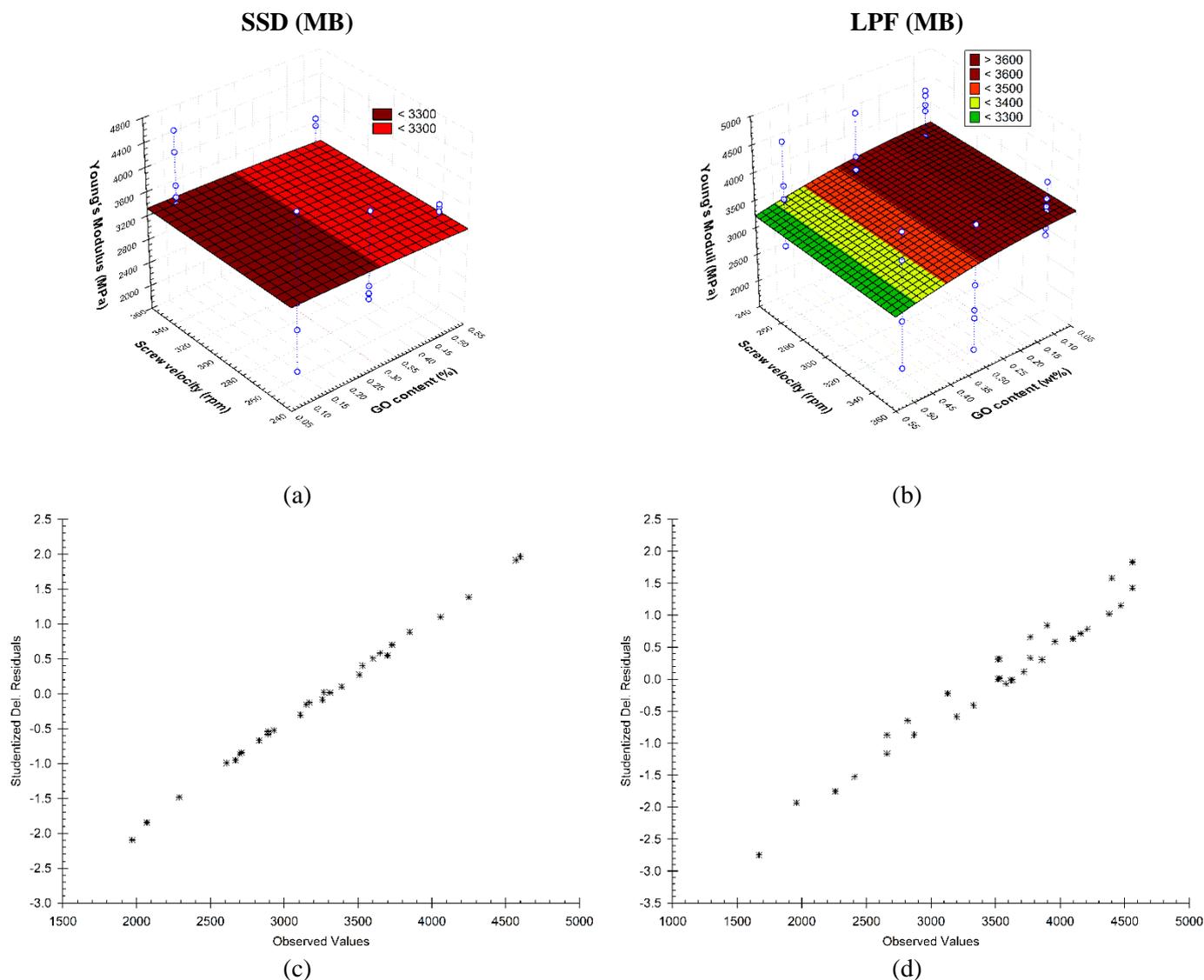

**FIGURE.S6** — Surface response for composites processing using MB dilution. (a) and (b): Surface response for Young's Moduli for SSD (MB) and LPF (MB), respectively; (c) and (d): Residual plots for Young's Moduli model of SSD (MB) and LPF(MB)

## SI.04 - Molecular Weight and morphology evaluation of PS and composites

Fig.S7 shows molecular weight distribution curve for neat and processed PS. A slight shift of the curve to lower molecular weight values when the polymer is processed can be seen, which is probably due to scission chain reactions that occurred during the processing. All molecular weight data (PS and composites) are presented on Tab.S6.

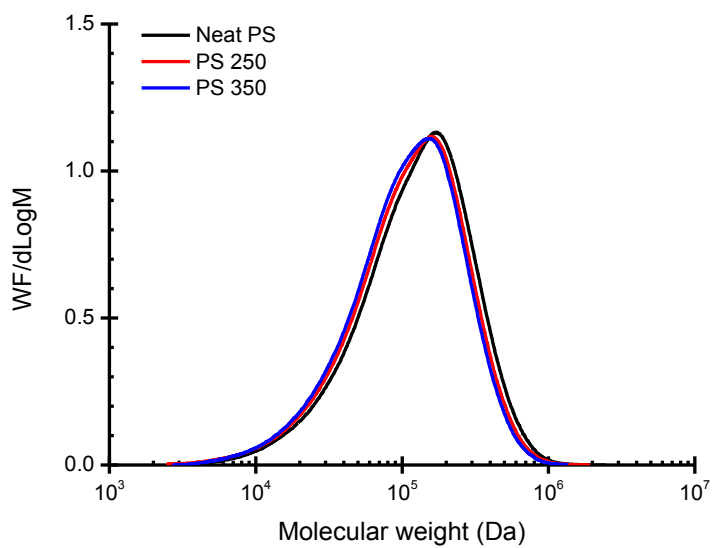

**FIGURE.S7** — Molecular weight distribution curves to neat PS and processed PS.

A Low Field Nuclear Magnetic Resonance (LF-NMR) study was developed to follow changes in $T_1H$ (lattice-spin relaxation time). In Fig.S8, a plot of $T_1H$ as function of molecular weight can be seen ($\overline{M}n$ and $\overline{M}w$) and there is no linear correlation between them.

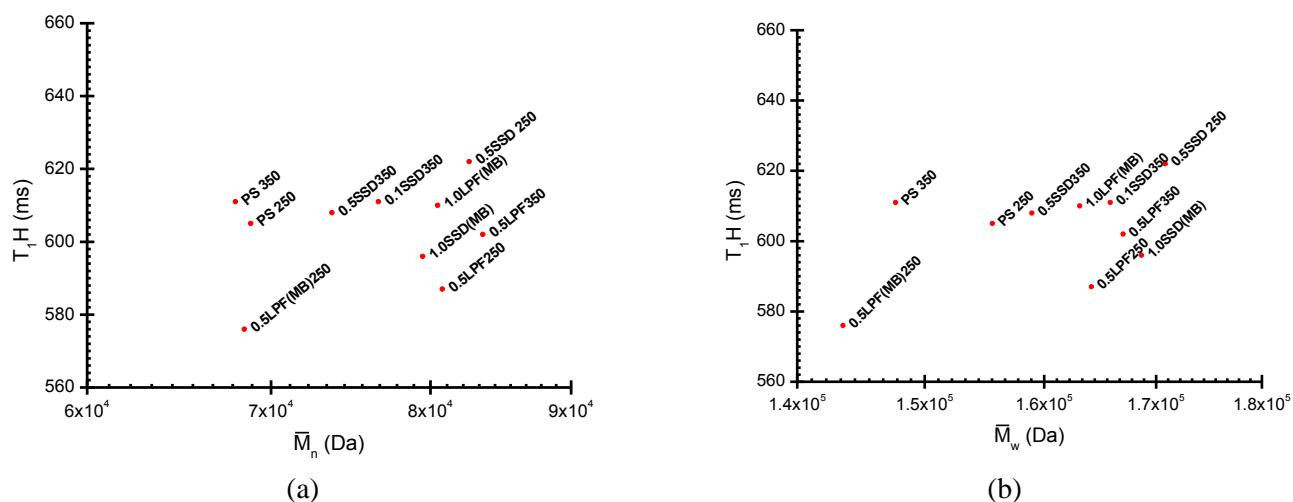

**FIGURE.S8** — Spin-lattice relaxation time as function of (a) $\overline{M}n$ and (b) $\overline{M}w$.

Tab.S6 is a résumé of Molecular weight, LF-NMR, and micro-tomography data. At least two volumes (analyzed volume) were chosen to calculate the object volume and object density for the aggregates. It is clear to notice that composites processed at 350 rpm present a tendency to generate a low quantity of aggregates with $T_1H$ values, slightly higher without any significant change in molecular weight. As discussed in the main text, this could be an indication of a better GO/polymer interface, consequently, a better mechanical performance is obtained. For the master-batches, there is a tendency for higher volumes of aggregates and lower $T_1H$ values, probably due to chain mobility in the vast majority of regions without the presence of the particles.

**TABLE.S6** — Tomography, LF-NMR and molecular weight data

| Sample | Analyzed volume (mm³) | Object volume (v%) | Object density (Obj/mm³) | $T_1H$ (ms) | $\overline{M}_n$ (Da) | $\overline{M}_w$ (Da) |
|---|---|---|---|---|---|---|
| **0.1SSD350** | 7.628 | 0.029 | 39.066 | 611 | 76584 | 165827 |
| **0.3SSD350** | 4.121 | 0.096 | 767.332 | — | — | — |
| **0.5SSD350** | 1.159 | 0.007 | 110.378 | 608 | 73668 | 158941 |
| **0.1SSD250** | 1.946 | 0.028 | 204.996 | — | — | — |
| **0.3SSD250** | | | | | | |
| **0.5SSD 250** | 1.519 | 0.070 | 472.517 | 622 | 82638 | 170840 |
| **0.1LPF350** | 1.329 | 0.034 | 112.885 | 601 | — | — |
| **0.5LPF350** | 0.713 | 0.038 | 384.189 | 602 | 83588 | 166994 |
| **0.5LPF250** | 1.201 | 0.044 | 310.523 | 587 | 80794 | 164154 |
| **1.0LPF(MB)** | 3.048 | 0.144 | 915.309 | 610 | 80485 | 163117 |
| **1.0SSD(MB)** | 0.825 | 0.099 | 691.144 | 596 | 79480 | 168661 |
| **0.5SSD(MB)250** | 1.558 | 0.039 | 236.865 | | — | — |
| **0.5SSD(MB)350** | 4.239 | 0.088 | 102.387 | | 72762 | 153310 |
| **0.5LPF(MB)250** | 1.413 | 0.076 | 1605.277 | 576 | 68447 | 143508 |
| **0.5LPF(MB)350** | 2.653 | 0.071 | 397.654 | | 72411 | 145929 |

## SI.05 - Different systems of polymer/2D material

Owing to the need to validate these processing methods, studies involving other polymer/2D composites are being developed. These studies allow the validation of this method for a different matrix - Poly(butylene adipate-co-terephthalate) – PBAT - and a different 2D material (Molybdenum disulfide - $MoS_2$). For PBAT/GO composites both methods were performed to create a 0.1 wt% GO composites, using GO suspended into ethanol (0.5 mg/mL) in order to reduce PBAT hydrolysis during processing. In this case, no changes were observed in the Young's Modulus of the two compared composites, however, about a 26% increase in strain at breaking point was observed for the composite obtained using the SSD method (Fig.S9a). This increase in SBP is an indication of a good dispersion of the particles and also a lubrication effect of GO inside the polymeric matrix caused by the sliding of the GO layer during composite plastic deformation. The SBP increase for a composite produced via LPF was lower than that produced by SSD, probably due to the PBAT degradation caused by the water content in ethanol. Another study is the PS/$MoS_2$ composite, and only the LPF method was tested since $MoS_2$ tends to re-stack during the rotovapor step of the SSD method. As well as GO, $MoS_2$ was dispersed into water (0.064 mg.mL$^{-1}$) and two formulations were processed: 0.05 wt% and 0.10 wt%. In Fig.S9b, stress-strain curves of PS/$MoS_2$ composites are presented. As can be observed, the presence of $MoS_2$ induces changes in the non-linear elastic region of curves increasing stress and strain at breaking point. The $MoS_2$ content has an effect on all measured mechanical properties even at a very low content. The 0.05 wt% composite showed higher UTS and SBP than 0.10 wt% composite.

All positive results presented for both systems are due to the good dispersion of the 2D particles in the polymer matrix indicating that the strategies used here, SSD and LPF, can be applied to different systems of polymer/2D materials in an industrially scalable polymer process with suitable morphology to obtain desirable properties.

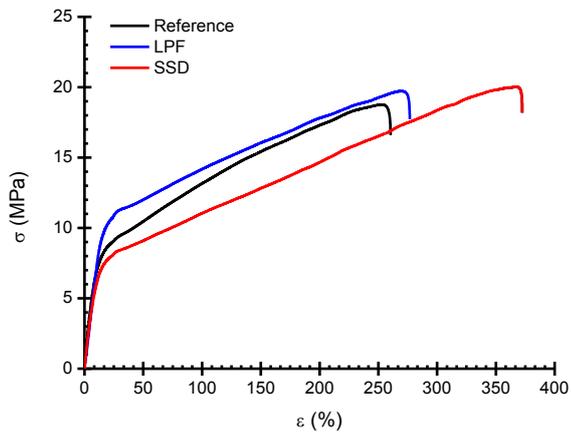 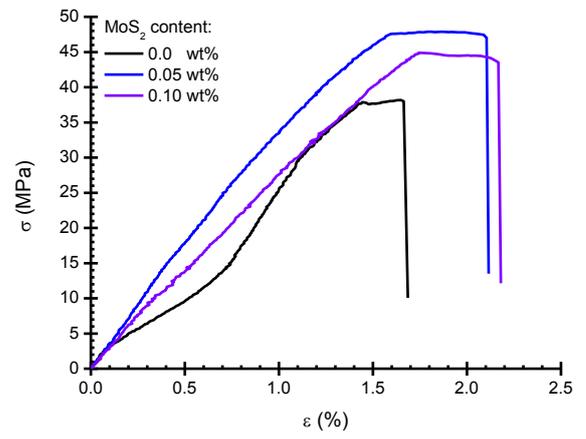

(a)            (b)

**FIGURE.S9** — Stress-strain curves. (a) PBAT/GO composites (0.1 wt%GO). (b) PS/MoS$_2$ composites (LFP).